\documentclass[11pt,a4paper,english,nofootinbib,superscriptaddress]{revtex4}
\usepackage{lmodern}

\usepackage[T1]{fontenc}
\usepackage[latin9]{inputenc}
\usepackage{babel}
\usepackage{color}
\usepackage{amsmath}
\usepackage{graphicx}
\usepackage{amssymb}
\usepackage{esint}
\usepackage[unicode=true, pdfusetitle,
 bookmarks=true,bookmarksnumbered=false,bookmarksopen=false,
 breaklinks=false,pdfborder={0 0 1},backref=false,colorlinks=false]
 {hyperref}
\usepackage{amsmath}
\usepackage{amssymb}
\def\b{\begin{equation}} \def\e{\end{equation}}
\def\bd{\begin{displaystyle}} \def\ed{\end{displaystyle}}
\def\ba{\begin{array}} \def\ea{\end{array}}

\def\bee{\begin{enumerate}}
\def\eee{\end{enumerate}}

\def\1{\mbox{I\hspace{-.15em}1}}

\def\R{{\rm I\hspace{-.15em}R}}

\def\C{\hspace{3pt}{\rm l\hspace{-.47em}C}}

\def\b{\begin{equation}}
\def\e{\end{equation}}
\def\bee{\begin{enumerate}}
\def\eee{\end{enumerate}}

\makeatletter
\usepackage{latexsym}\usepackage{bm}

\makeatother
\begin{document}

\title{de Sitter super-gravity in ambient space formalism}

\author{M.V. Takook}
\email{takook@razi.ac.ir} 
\affiliation{Department of Physics, Razi University,
Kermanshah, Iran}
\affiliation{Department of Physics,
Science and Research branch, \\Islamic Azad University, Tehran,
Iran}

\date{\today}

\begin{abstract}
In the de Sitter ambient space formalism the massless fields, which include the linear gravity and massless minimally coupled scalar field, can be written in terms of two separate parts: a massless conformally coupled scalar field and a polarization tensor(-spinor) part. Therefore due to the massless conformally coupled scalar field, there exist an unique Bunch-Davies vacuum state for quantum field theory in the de Sitter space time. In the de Sitter ambient space formalism one can show that the massless fields with spin $s\geq 1$ are gauge invariant. By coupling the massless gauge spin-$2$ and the massless gauge spin-$3/2$ fields and using  the super-symmetry algebra in de Sitter ambient space formalism, one can naturally construct a unitary de Sitter super-gravity on Bunch-Davies vacuum state.
\end{abstract}

\maketitle

\section{Introduction}

The construction of the quantum field theory (QFT) in the de Sitter (dS) universe, poses some serious problems. The first problem is that the dS gravitational field propagator in the usual linear approximation exhibits a large-distance pathological behaviour, {\it i.e.} an infra-red divergence \cite{al,flilto,anmo}. Iliopoulos et al. have shown that the infra-red divergence of the graviton propagator on dS background does not manifest itself in the quadratic part of the effective action in the one-loop approximation \cite{anilto}. This means that the pathological behaviour of the graviton propagator may be gauge dependent and so should not appear in an effective way as a physical quantity.

The second problem is the non-existence of super-gravity models in
dS space-time. Such arguments are often based on the non-existence of Majorana spinors for O(1, 4) \cite{pivaso,dehe}. Pilch et al. have shown that the dS super algebra can be closed only for even $N$ and if for every spinor, its independent charge-conjugate could be defined, dS super-gravity can be established with even $N$ but the action becomes either imaginary or contains vector-ghosts \cite{pivaso,anfrma}.  Recently, Freedman et al. have presented an exemplary review of dS super-gravity and its problems \cite{anfrma}.

In this letter by the use of the gauge theory in the de Sitter ambient space formalism, which contains the main results of a previous paper \cite{ta1403}, we have properly addressed all the above mentioned problems. The ambient space formalism allows us to reformulate the QFT in a rigorous mathematical framework, based on the analyticity of the complexified pseudo-Riemannian manifold and the group representation theory. By using the spinor representation of the universal covering group of dS group, $Sp(2,2)$, in the ambient space formalism, and its charge conjugation, a novel dS super-algebra for odd $N$ can be obtained \cite{parota}. In this formalism the QFT in dS universe,  which include the gravitational field, can be constructed on Bunch-Davies vacuum state. Then one can construct the dS super-gravity in this formalism by the gauge principle.

\section{Bunch-Davies vacuum state}

 The dS ambient space formalism is visualized as a $4$-dimensional hyperboloid embedded in a $5$-dimensional Minkowski space-time with equation:
\begin{equation} \label{dSs} M_H=\left\lbrace x \in \R^5 |\; \; x \cdot x=\eta_{\alpha\beta} x^\alpha
x^\beta =-H^{-2}\right\rbrace,
\end{equation} 
where $\eta_{\alpha\beta}=$diag$(1,-1,-1,-1,-1)$, $\alpha,\beta=0,1,2,3,4,$ and $H$ is constant Hubble parameter. The dS metrics is 
\b ds^2=\eta_{\alpha\beta}dx^{\alpha}dx^{\beta}|_{x^2=-H^{-2}}=g_{\mu\nu}^{dS}dX^{\mu}dX^{\nu},\e
with $X^\mu$ as a 4 space-time intrinsic coordinates on dS hyperboloid ($\mu=0,1,2,3$). The tensor or tensor-spinor fields on dS hyperboloid in the ambient space formalism satisfy the transversality condition: $ x^\alpha {\cal K}_\alpha(x)=0$ and they are also homogeneous functions with degree $\lambda$: $x\cdot\partial {\cal K}_\alpha(x)=\lambda {\cal K}_\alpha(x).$ The projection operator on dS hyperboloid is defined as:
$\theta_{\alpha\beta}=\eta_{\alpha\beta}+H^2x_\alpha x_\beta$, $x.\theta=0$.

The unitary irreducible representations (UIR) of dS group was completed by Takahashi \cite{tak}. The analyticity in complexified dS space-time had been studied by Bros et al \cite{brgamo,brmo}. By combining these two subjects, one can construct the quantum field operators, the quantum states and the two point functions for the various spin fields in dS space-time on a unique vacuum state {\it i.e.} the Bunch-Davies vacuum state. The massive fields in the dS space-time correspond to the principal series representation of the dS group and propagate inside the dS light-cone \cite{brgamo,gata,gagata,taazba,berotata,ta1403}. They correspond to the massive Poincar\'e fields in the null curvature limit. For constructing the quantum field operator, one must first define the creation and annihilation operators on corresponding Hilbert space ${\cal H}$. The Hilbert space and the quantum states can be exactly defined by the UIR of the principal series \cite{tak,ta1403}. The massive field operators are defined as a map on the Fock space:
\begin{equation} \label{fockspacem}
\mbox{Massive Field Operators}: \;{\cal F}({\cal H}) \longrightarrow {\cal F}({\cal H}),
\end{equation} 
which is constructed by the corresponding Hilbert spaces. The massive tensor(-spinor) field can be written in terms of the massive scalar field \cite{brgamo,gata,gagata,taazba,berotata,ta1403}:
\begin{equation}
{\cal K}_{\alpha_1...\alpha_l}^{(\nu)}(x)=D_{\alpha_1...\alpha_l}(x,\partial,\nu) \phi^{(\nu)}(x),
\end{equation}
where $\nu$ is the principal series parameter. 

The quantum massive scalar field, $\phi^{(\nu)}(x)$, in ambient space formalism was constructed previously 
in terms of the dS plane wave $(x.\xi)^{\lambda}$ with $\lambda \in \C$ and $\Im \lambda<0$ \cite{brgamo,brmo}. The $\xi^\alpha$ is a null five-vector in the positive cone $C^+=\left\lbrace \xi \in \R^5|\;\; \xi\cdot \xi=0,\;\; \xi^{0}>0 \right\rbrace.$ In the null curvature limit it becomes the energy-momentum four-vector $k^\mu=(k^0, \vec{k})$. The homogeneous degree $\lambda$ in the null curvature limit has a relation with the mass in Minkowski space-time. 
For obtaining a well-defined field operator, it must be constructed on the complex dS space-time \cite{brgamo,brmo}, $$ M_H^{(c)}=\{ z=x+iy\in \C^5|\;\eta_{\alpha \beta}z^\alpha z^\beta=-H^{-2}\}.$$ The quantum field operator in this notation can be written in the following form \cite{brgamo,brmo,ta1403}:
$$ \phi^\nu(z)=\int d\mu(\xi) \left[ a({\bf\tilde{\xi}},\nu)(z\cdot\xi)^{-\frac{3}{2}-i\nu}
+ a^\dag(\xi,\nu)(z\cdot\xi)^{-\frac{3}{2}+i\nu}
\right] ,$$
where $\tilde \xi^\alpha=(1, -\vec \xi, \xi^4)$ and $d\mu(\xi)$ is dS invariant volume element on three-sphere $S^3$ \cite{ta1403}.  The vacuum state and the ''one-particle'' state are defined as, $a(\xi,\nu)|\Omega>=0, a^\dag(\xi,\nu)|\Omega>=N|\xi,m_j;j=0,p=-\frac{1}{2}+i\nu>$. $N$ is the normalization constant, $j$ and $p$ classify the UIR of de Sitter group and in the null curvature limit play the role of the spin and the mass respectively. The quantum states $|\xi,m_j;j=0,p=-\frac{1}{2}+i\nu>$ is defined explicitly by the UIR of dS group \cite{tak,ta1403}.
The analytic two-point function is then calculated directly from the complex dS plane wave and the vacuum states, up to the normalization constant, $$W^{(\nu)}(z,z')=\;<\Omega|\phi^\nu(z)\phi^\nu(z')|\Omega>=2\pi^2e^{\pi \nu} H^3 c_{\nu} P^{(5)}_{-\frac{3}{2}+i\nu}(H^2z.z') ,$$ 
where $P^{(5)}_{-\frac{3}{2}+i\nu}(H^2z.z')$ is the  generalized Legendre function of the first kind \cite{brmo}. The normalization constant $c_{\nu}$ is fixed by imposing the Hadamard condition, which fix the vacuum state as the Bunch Davis vacuum state \cite{brmo}. Thus the vacuum state is fixed leading to 
$ c_{\nu}=\frac{e^{-\pi
\nu}\Gamma(\frac{3}{2}+i\nu)\Gamma(\frac{3}{2}-i\nu)}{2^5\pi^4H}$  \cite{brmo}.
The two point function ${\cal W}^{(\nu)}(x,x')$ is the "boundary value" (in a distributional sense) of the analytic function $W^{(\nu)}(z,z')$. 
The two-point function for various massive spin fields can be written in terms of polarization bi-tensor(-spinor) parts and two point functions of the massive scalar field \cite{gata,bagamota,gagata,taazba,ta1403}
\begin{equation}
{\cal W}^{(\nu)}_{\alpha_1..\alpha'_1..}(x,x')={\cal D}_{\alpha_1..\alpha'_1..}(x,\partial;x',\partial';\nu) {\cal W}^{(\nu)}(x,x').
\end{equation}

\section{Massless Fields}

The massless conformally coupled scalar field corresponds to the complementary series representation of the dS group \cite{brmo}. This field can be obtained by replacing the parameter $\nu$ in the principal series representation for massive scalar field by $\nu=-\frac{i}{2}$:
$$ \phi_c(z)=\int d\mu(\xi) \left[ a\left({\bf\tilde{\xi}},-\frac{i}{2}\right)(z\cdot\xi)^{-2}
+ a^\dag\left(\xi,-\frac{i}{2}\right)(z\cdot\xi)^{-1}
\right] .$$
The analytic two-point function for this field is \cite{brgamo,brmo}:
$$ W_c(z,z')=\frac{-iH^2}{2^4\pi^2} P_{-1}^{(5)}(H^2 z\cdot z')=
\frac{H^2}{8\pi^2}\frac{-1}{1-{\cal Z}(z,z')}. $$
The vacuum state in this case is also the Bunch-Davies vacuum state. The Wightman two-point function ${\cal W}_c(x,x')$ is the "boundary value" of the function $W_c(z, z')$ \cite{brmo}:
$$ {\cal W}_c(x,x')=\frac{-H^2}{8\pi^2}\lim_{\tau \rightarrow 0}\frac{1}{1-{\cal
Z}(x,x')+i\tau\epsilon(x^0,x'^0)}$$ \begin{equation}\label{stpci2}
=\frac{-H^2}{8\pi^2}\left[
P\frac{1}{1-{\cal Z}(x,x')}
-i\pi\epsilon(x^0,x'^0)\delta(1-{\cal Z}(x,x'))\right],\end{equation}
where the symbol $P$ represents the principal part. ${\cal Z}$ is the geodesic distance between two points $x$ and $x'$ on the dS hyperboloid, 
$${\cal Z}(x,x')=-H^2 x\cdot x'=1+\frac{H^2}{2} (x-x')^2,  \;\;  \epsilon (x^0-x'^0)=\left\{\begin{array}{clcr} 1&\;\;x^0>x'^0 ,\\ 0&\;\;x^0=x'^0,\\ -1&\;\;x^0<x'^0 .\\ \end{array} \right. $$
A massless spinor field corresponds to the discrete series representation of dS group which is conformally invariant \cite{ta1403,bagamota} and can be written in terms of the massless conformally coupled scalar field \cite{ta1403}, $$ \psi(x)= \left(-\not x \not \partial^\top+1\right) \phi_c(x),$$ where $\partial^\top_\alpha\equiv\theta_{\alpha\beta}\partial^\beta=\partial_\alpha+H^2x_\alpha
x\cdot\partial$ and $\not x=x^\alpha \gamma_\alpha$. There are $5$ $\gamma_\alpha$ matrices where $\gamma_\alpha^\dag=\gamma_0\gamma_\alpha\gamma_0$ and $\{\gamma^{\alpha},\gamma^{\beta}\}=2\eta^{\alpha\beta}$ \cite{bagamota}. These two massless field operators, transform by the UIR of the dS group, and can be calculated explicitly, similar to the massive cases  by defining the creation and annihilation operators on the corresponding Hilbert spaces \cite{ta1403}. By defining the creation and annihilation operators on the corresponding Hilbert spaces, the quantum free field operators can be explicitly calculated similar to the massive cases \cite{ta1403}. 

A massless minimally coupled scalar field can be written in terms of the massless conformally coupled scalar field in the dS ambient space formalism by using the identity \cite{ta1403}:
\begin{equation} \label{msfincsf} \phi_m(x)= \left[A\cdot\partial^\top + 2 A\cdot x\right]\phi_c(x), \end{equation} 
  where $A^\alpha$ is an arbitrary constant five-vector which can be fixed in the null curvature limit in the interaction cases \cite{taga17}. The two-point function can be defined on the vacuum state of the conformally coupled scalar field or Bunch-Davies vacuum state as:
$$ {\cal W}_{m}=\left[A\cdot\partial^\top + 2 A\cdot x\right]\left[A\cdot\partial'^\top + 2 A\cdot x'\right]{\cal W}_{c},$$ 
It should be noted that this two-point function is analytic and free of any infra-red divergence.

For the massless fields with spin=$j=p\geq 3$, which associate to the discrete series representations, the homogeneous degrees of tensor(-spinor) fields contain real and positive number ($\lambda_1\geq 0 )$ \cite{ta1403}, therefore the dS plane wave cannot be defined properly since the plane wave solution has singularity in $x\longrightarrow \infty $. Therefore these fields cannot propagate on dS space.

For massless fields with spin=$j=p=1,\frac{3}{2}$ and $2$, the Hilbert spaces and consequently the quantum states cannot be defined uniquely and it appear a gauge invariance \cite{tak,ta1403}, which will be recalled in the next section. There exist the three types of massless gauge fields on the dS ambient space formalism; spin-$1$ vector fields $K_\alpha$, spin-$\frac{3}{2} $ vector-spinor fields $\Psi_\alpha$ and spin-$2$ fields. For a spin-$2$ field there are two possibilities for construction of the quantum field operator: a rank-$2$ symmetric tensor field $H_{\alpha\beta}$ and a rank-$3$ mix-symmetric tensor field ${\cal K}_{\alpha\beta\gamma}$ \cite{fr84}, $${\cal K}_{\alpha\beta\gamma}=-{\cal K}_{\alpha\gamma\beta},\; \;{\cal K}_{\alpha\beta\gamma}+{\cal K}_{\gamma\alpha\beta}+{\cal K}_{\beta\gamma\alpha}=0.$$ The rank-$2$ symmetric tensor field, in dS ambient space formalism, can be written in terms of massless minimally coupled scalar field \cite{taro12,khrota}, but breaks the conformal transformation \cite{bifrhe,tatafa,tata}. For preserving the
conformal transformation, the spin-$2$ field may be represented by a rank-$3$ mix-symmetric tensor field ${\cal K}_{\alpha\beta\gamma}$ \cite{bifrhe,tatafa,tata}. The tensor field ${\cal K}_{\alpha\beta\gamma}$ can be written in terms of tensor field $H_{\alpha\beta}$ by imposing subsidiary conditions \cite{fr84,tatafa}. In order to properly define massless quantum fields and the quantum states, one must first fix the gauge. Since the action of the creation operator on the Hilbert space results in states which are out of the Hilbert space. Then a new vector spaces are constructed by using the representation of dS group as follows:
$$ {\cal M}=V_1 \oplus V_2 \oplus V_3, $$
where $V_1 $ and $V_3$ are the space of the gauge dependent states and the pure gauge states, respectively. The space $V_2/V_3\equiv {\cal H}$ is a vector space containing the physical states which are the Hilbert spaces constructed by the corresponding UIR of discrete series. This is known as the Gupta-Bleuler triplets \cite{anla,la,gareta}. In these cases the massless field operators are defined as a map on the Fock space which are constructed on the vector space ${\cal M}$:
$$ \mbox{Massless Field Operators}: \; {\cal F}({\cal M}) \longrightarrow {\cal F}({\cal M}).$$

The structure of un-physical states $V_3$ and $V_1$ are obtained by using the gauge transformation and the gauge fixing field equation, respectively. The gauge transformation and the gauge invariant field equations of the vector field, the vector-spinor field, the spin-$2$ rank-$2$ symmetric tensor field and the spin-$2$ rank-$3$ mix-symmetric tensor field have been presented in the previous papers in terms of the Casimir operator of dS group \cite{ta1403,gagarota,azam,derotata,taro12} which will be also recalled in the next section. 

In ambient space formalism, the  two-point function of free massless gauge quantum fields can be written in terms of the two-point  function of massless conformally coupled scalar fields: \cite{gagarota,bagamota,ta1403}
$$ {\cal W}_{\alpha_1..\alpha'_1..}(x,x')={\cal D}_{\alpha_1..\alpha'_1..}(x,\partial;x',\partial'){\cal W}_c(x,x') .$$ Thus the two point function of linear gravity is free of any infrared divergence and therefore the infrared divergence problem is exactly solved in this formalism \cite{ta1403,khrota}. 

\section{Gauge invariant field equation}

\subsection{In Hilbert space}

First we present the field equation for the quantum states  $\left.|\xi;j,p\right>\equiv\left.|{\bf q};j,p\right>$. ${\bf q}$ is a quaternion with norm $|{\bf q}|=r=\sqrt{q_1^2+q_2^2+q_3^2+q_4^2}<1$ and $\xi^\alpha\equiv  (1,\frac{{\bf q}}{|{\bf q}|})$. The partial differential equation for the quantum state $\left.|{\bf q};j,p\right>$,
 which  is obtain from the Casimir operator, is \cite{tak,ta1403}: 
\b\label{difequ} \left[\frac{1}{4}(1-|{\bf q}|^2)\triangle -p{\cal D}-\frac{1}{2}(A_j{\cal D}_1+B_j{\cal D}_2+C_j{\cal D}_3)+(j(j+1)-p(p+1))\right]\left|{\bf q};j,p\right>=0,\e
where the differential operators $\triangle, {\cal D}, {\cal D}_1,{\cal D}_2$ and ${\cal D}_3$ are defined explicitly in ${\bf q}$-space by Takahashi \citep{tak}: 
$$ \triangle=\frac{\partial^2}{\partial q_1^2}+\frac{\partial^2}{\partial q_2^2}+\frac{\partial^2}{\partial q_3^2}+\frac{\partial^2}{\partial q_4^2}=\frac{\partial^2}{\partial r^2}+\frac{3}{r}\frac{\partial}{\partial r}+\frac{1}{r^2}\triangle_{S^3},$$
$${\cal D}=q_1\frac{\partial}{\partial q_1}+q_2\frac{\partial}{\partial q_2}+q_3\frac{\partial}{\partial q_3}+q_4\frac{\partial}{\partial q_4}=r\frac{\partial}{\partial r}+{\bf u}\cdot \frac{\partial}{\partial {\bf u}},$$
$${\cal D}_1=q_1\frac{\partial}{\partial q_2}-q_2\frac{\partial}{\partial q_1}+q_4\frac{\partial}{\partial q_3}-q_3\frac{\partial}{\partial q_4}=u_1\frac{\partial}{\partial u_2}-u_2\frac{\partial}{\partial u_1}+u_4\frac{\partial}{\partial u_3}-u_3\frac{\partial}{\partial u_4},$$
$${\cal D}_2=q_1\frac{\partial}{\partial q_3}-q_3\frac{\partial}{\partial q_1}+q_2\frac{\partial}{\partial q_4}-q_4\frac{\partial}{\partial q_2}=u_1\frac{\partial}{\partial u_3}-u_3\frac{\partial}{\partial u_1}+u_2\frac{\partial}{\partial u_4}-u_4\frac{\partial}{\partial u_2},$$
\b\label{relations} {\cal D}_3=q_1\frac{\partial}{\partial q_4}-q_4\frac{\partial}{\partial q_1}+q_3\frac{\partial}{\partial q_2}-q_2\frac{\partial}{\partial q_3}=u_1\frac{\partial}{\partial u_4}-u_4\frac{\partial}{\partial u_1}+u_3\frac{\partial}{\partial u_2}-u_2\frac{\partial}{\partial u_3}.\e
 $\triangle_{S^3} $ is the Laplace operator on the hypersphere $S^3$. The $A_j,\; B_j$ and $C_j$ are $(2j+1)\times (2j+1)$-matrix representations of the generators of the $SU(2)$ group \cite{tak}. ${\bf u}$ is a quaternion with norm $1$ (${\bf q}=r{\bf u}$).
 
The case $j=p\geq 1$ is of particular interest, since it corresponds to the massless fields in the dS space with existing equivalent massless fields in the Minkowski space-time in the null curvature limit. The field equation (\ref{difequ}) for these cases ($j=p\geq 1$) become:
\b\label{difequj=p} \left[\frac{1}{4}(1-|{\bf q}|^2)\triangle -j{\cal D}-\frac{1}{2}(A_j{\cal D}_1+B_j{\cal D}_2+C_j{\cal D}_3)\right]\left|{\bf q};j,j\right>=0.\e 
Since $\triangle, {\cal D}, {\cal D}_1,{\cal D}_2$ and ${\cal D}_3$ are differential operators, the quantum state $|{\bf q};j,j>$ cannot be uniquely defined by this equation, and also, it is invariant under the following transformation:
\b \label{gatrq} \left|{\bf q};j,j\right> \Longrightarrow \left|{\bf q};j,j\right>^g= \left|{\bf q};j,j\right>+\left|{\bf q}_0;j,j\right>,\e
where  ${\bf q}_0$ is a constant quaternion.

\subsection{In space time}
Here, we only consider three cases, namely $j=p= 1, \frac{3}{2}$ and $ 2$, which correspond to the massless vector, vector-spinor and spin-$2$ fields. The vector-gauge invariant field equation can be rewritten in the following form \cite{gagarota}:
$$ Q^{(1)}_1 K_\alpha^{\;\;a}+ \nabla^\top_\alpha \partial\cdot K^{\;a}=0.$$
This field equation is invariant under the following gauge transformation:
$$ K_\alpha^{\;\;a} \longrightarrow (K^g)_\alpha^{\;\;a}=K_\alpha^{\;\;a} +\nabla^\top_\alpha \phi^a,$$
where $\phi^a$ is an arbitrary scalar fields.  $  Q^{(1)}_1$ is the Casimir operator act on a vector field \cite{ta1403} and the transverse-covariant derivative which acts on tensor field is \cite{tata}:
\b \label{dscdrt}
\nabla^\top_\beta T_{\alpha_1....\alpha_l}\equiv 
\partial^\top_\beta
T_{\alpha_1....\alpha_l}-H^2\sum_{n=1}^{l}x_{\alpha_n}T_{\alpha_1..\alpha_{n-1}\beta\alpha_{n+1}..\alpha_l}.\e 

The gauge invariant field equation for vector-spinor field is \cite{paenta}
\b \label{gifes32} \left(Q_{\frac{3}{2}}^{(1)}+\frac{5}{2}\right) \Psi_\alpha+ \nabla^\top_{ \alpha}\partial^\top\cdot\Psi=0,\e
which is invariant under the gauge transformation \cite{azam,fatata}:
$$\Psi_\alpha \longrightarrow \Psi^g_\alpha=\Psi_\alpha+\nabla^\top_{\alpha } \psi.$$
$\psi$ is an arbitrary spinor field and $  Q_{\frac{3}{2}}^{(1)}$ is the Casimir operator act on a vector-spinor field \cite{ta1403}. The transverse-covariant derivative which acts on tensor-spinor field is defined by the following relation:
\b \label{cdsa} \nabla^\top_\beta \Psi_{\alpha_1....\alpha_l}\equiv 
\left(\partial_\beta^\top+\gamma^\top_\beta\not x\right)
\Psi_{\alpha_1....\alpha_l}-H^2\sum_{n=1}^{l}x_{\alpha_n}\Psi_{\alpha_1..\alpha_{n-1}\beta\alpha_{n+1}..\alpha_l}.\e  

The gauge invariant field equation for spin-$2$ rank-$2$ symmetric tensor field ${\cal H}_{\alpha\beta}={\cal H}_{\beta\alpha}$ is \cite{taro12}:
\b \label{festgi} \left( Q_2^{(1)}+6\right){\cal H}(x)+D^\top_2\partial_2^\top\cdot {\cal H}(x)=0,\e
which is invariant under the following gauge transformation:
\b \label{gfst3} {\cal H}_{\alpha\beta}\Longrightarrow {\cal H}_{\alpha\beta}^g={\cal H}_{\alpha\beta}+\left(D_2^\top A\right)_{\alpha\beta}. \e   
$A_\beta(x)$ is an arbitrary vector field and $  Q_{2}^{(1)}$ is the Casimir operator act on a rank-$2$ symmetric tensor field \cite{ta1403}. The generalized gradient $D_2^\top$ is defined by: $$\left(D_2^\top A\right)_{\alpha\beta}=\nabla^\top_\alpha A_\beta+\nabla^\top_\beta A_\alpha.$$ $\partial_2^\top\cdot {\cal H}$ is the generalized divergence: $ \partial_2\cdot
 {\cal H}=\partial^\top\cdot {\cal H}-\frac{1}{2}H^2D_1 {\cal H}'=\partial \cdot {\cal H}- H^2 x
 {\cal H}'-\frac{1}{2} \bar  \partial {\cal H}',$ and ${\cal H}'={\cal H}_\alpha^\alpha$.
 
 Using the dS algebra, one can obtain the following gauge invariant field equation for spin-$2$ rank-$3$ mixed symmetry tensor field \cite{ta1403}:
\b \label{ran3mieq} \left(Q^{(1)}_3+6\right) {\cal K}_{\alpha\beta\gamma}+  \nabla^\top_\alpha \left( \partial_3^\top\cdot {\cal K}\right)_{\beta\gamma}=0,\e
which is invariant under the gauge transformation:
\b \label{gts2r3} {\cal K}_{\alpha \beta\gamma} \longrightarrow {\cal K}_{\alpha \beta\gamma}^g={\cal K}_{\alpha \beta\gamma}+ \nabla^\top_\alpha \Lambda_{\beta\gamma}.\e
$\Lambda_{\alpha\beta}$ is an arbitrary rank-$2$ anti-symmetric tensor field ($x\cdot \Lambda_{.\alpha}=0, \; \Lambda_{\alpha\beta}=-\Lambda_{\beta\alpha}$). The traceless condition on ${\cal K}$ results in the divergence-less condition on the pure gauge field $\Lambda$, $\partial^\top \cdot \Lambda_{.\alpha}=0$. The generalized divergence $\partial_3^\top\cdot$ is defined by:
$$  \left( \partial_3^\top\cdot {\cal K}\right)_{\beta\gamma} \equiv \nabla^\top\cdot {\cal K}_{.\beta\gamma}+\nabla^\top\cdot {\cal K}_{\beta.\gamma} ,$$
with the condition $\nabla^\top\cdot {\cal K}_{\beta.\gamma}=-\nabla^\top\cdot {\cal K}_{\gamma.\beta} $. $  Q_{3}^{(1)}$ is the Casimir operator act on a  rank-$3$ mixed symmetry tensor field \cite{ta1403}.

\section{Gauge Theory}

One can associate a local symmetrical group (or super-group) to these gauge transformations. The gauge invariant field equations and the gauge transformations can also be obtained by using the gauge principle \cite{ta1403}. The Lagrangian is defined through the gauge-covariant derivative which, in turn, defines a quantity that preserves the gauge invariant transformation of the Lagrangian. 

The massless gauge vector fields $K_\alpha$ can be associated to the electromagnetic, weak and strong nuclear forces in the frame work of the abelian and non-abelian gauge theories \cite{ta1403}. 
Now we attempt to reformulate this gauge theory in ambient space formalism. The vector gauge potentials $K_\alpha^{\;\;a}$ with $a=1,2,..., n^2-1$, can be associated with the gauge group $SU(n)$. One can assume that $t^a$ is the generator of $SU(n)$ group, satisfying the following commutation relation
\b  \label{yma}[t_a,t_b]=C^{\;\;\;\;c}_{ab}t_c.\e
The notation of local gauge symmetry with its space-time-dependent transformation can be used to generate the gauge interaction. 
For obtaining a local gauge invariant Lagrangian, it is
necessary to replace the transverse-covariant derivative
$ \nabla^\top_\beta$ (\ref{dscdrt}) with the gauge-covariant derivative
$D_\beta^{K}$ which is defined by 
\b D_\beta^K \equiv\nabla^\top_\beta +iK_{\beta}^{\;\;a}t_a, \e
where the gauge potential or connection $K_{\beta}^{\;\;a}$ is a massless vector field. Considering a local infinitesimal gauge transformation generated by $\epsilon^a(x)t_a$, one has
$$ \delta_{\epsilon}K_\alpha^{\;\;a}=D^K_\alpha\epsilon^a=\partial^\top_\alpha \epsilon^a+ C_{cb}^{\;\;\;\;a}K_\alpha^{\;\;c}\epsilon^b.$$
The ''curvature'' ${\cal R}$, for the Lie algebra of the gauge group $SU(N)$, is defined by:
$${\cal R}(D^K_\alpha,D^K_\beta)=-[D^K_\alpha,D^K_\beta]=R_{\alpha\beta}^{\;\;\;\;a}t_a,$$
where
$$ R_{\alpha\beta}^{\;\;\;\;a}=\nabla^\top_\alpha K_\beta^{\;\;a}-\nabla^\top_\beta K_\alpha^{\;\;a}+K_\beta^{\;\;b}K_\alpha^{\;\;c}C_{bc}^{\;\;\;\;a},\;\;\; \;\; x^\alpha R_{\alpha\beta}^{\;\;\;\;a}=0= x^\beta R_{\alpha\beta}^{\;\;\;\;a}.$$

The $SU(n)$ gauge invariant action or Lagrangian in the dS background for the gauge field $K_\alpha^{\;\;a}$ is:
$$ S[K]=-\frac{1}{2}\int d\mu(x) tr \left({\bf R}_{\alpha\beta}{\bf R}^{\alpha\beta} \right)=-\frac{1}{2}\int d\mu(x) R_{\alpha\beta}^{\;\;\;\;a}R^{\alpha\beta b} tr\left(t_at_b\right) , $$ 
with ${\bf R}_{\alpha\beta}= R_{\alpha\beta}^{\;\;\;\;a}t_a$ and summing over the repeated indices  and $d\mu(x)$ is the dS invariant volume element \cite{brmo}. The normalization of the structure constant is usually
fixed by requiring that, in the fundamental representation, the corresponding
matrices of the generators $t_a$ are normalised such as
$  tr\left(t_at_b\right) =\frac{1}{2} \delta_{ab}.$

The above construction can be simply generalized to the de Sitter group $SO(1,4)$  with the $10$ generators $L_{\alpha\beta}$ and the following commutation relation \cite{niobso}:
$$
[L_{\alpha\beta}, L_{\gamma\delta}] =
-i(\eta_{\alpha\gamma}L_{\beta\delta}+\eta_{\beta\delta}
L_{\alpha\gamma}-\eta_{\alpha\delta}L_{\beta\gamma}-\eta_{\beta\gamma}
L_{\alpha\delta})\Longrightarrow  [t_a,t_b]=C^{\;\;\;\;c}_{ab}t_c, a,b,c=1,...,10.$$
In this gauge gravity model we have 10 vector field  $K_\alpha^{\;\; a}$ with the conditions $x^\alpha K_\alpha^{\;\; a}=0$ and $40$ degrees of freedom.  By using these $10$ vector field one can construct a rank-$3$ tensor field with the following relations:
\b K_\alpha^{\;\; a}=X^{a\beta\gamma}{\cal K}_{\alpha\beta\gamma}, \;\;{\cal K}_{\alpha\beta\gamma}=-{\cal K}_{\alpha\gamma\beta}, \;\; x^\alpha {\cal K}_{\alpha\gamma\beta}=0.\e
The projection of this tensor field on de Sitter hyperboloid, $x^\beta {\cal K}_{\alpha\beta\gamma}=0=x^\gamma {\cal K}_{\alpha\beta\gamma}$, results to a tensor field with $24$ degrees of freedom. We obtain the $20$ degrees of freedom with the subsidiary condition ${\cal K}_{\alpha\beta\gamma}+{\cal K}_{\beta\gamma\alpha}+{\cal K}_{\gamma\alpha\beta}=0$. Now we have a rank-$3$ mixed symmetry tensor field which can be associated with a spin-$2$ field. The physical state of the vector field satisfy the following field equation and divergence-less condition \cite{gagarota}:
$$ \left(Q_{0}^{(1)}-2\right)K_\alpha^{\;\;a}=0, \;\; \partial^\top \cdot K^a=0, $$ 
then we obtain: 
$$ \left(Q_{0}^{(1)}-2\right)X^{a\beta\gamma}{\cal K}_{\alpha\beta\gamma}=0,\;\; {\cal K}_{.\beta\gamma}\cdot\partial^\top X^{a\beta\gamma}+X^{a\beta\gamma}\partial^\top \cdot {\cal K}_{.\beta\gamma}=0 .$$
 $  Q^{(1)}_0$ is the Casimir operator act on a scalar field, $Q^{(1)}_0=-H^{-2}\partial^\top\cdot\partial^\top=-H^{-2}\square_H,$
where $\square_H$ is the Laplace-Beltrami operator on dS space-time. By imposing the following conditions on the coefficient $X^{a\beta\gamma}$:
$$ (Q_{0}-2)X^{a\beta\gamma}=0,\;\;\partial^{\top\alpha}X^{a\beta\gamma} \partial^{\top}_{\alpha}{\cal K}_{\rho\beta\gamma}=0,\;\;   {\cal K}_{\alpha\beta\gamma}\partial^{\top\alpha}X^{a\beta\gamma}=0,$$
and the divergence-less condition $\partial^\top \cdot {\cal K}_{.\beta\gamma}=0=\partial^\top \cdot {\cal K}_{\alpha.\beta}$, the tensor field ${\cal K}_{\alpha\beta\gamma} $ satisfy the following field equation
$$ Q_{0}^{(1)}{\cal K}_{\alpha\beta\gamma}=0. $$ 
If we impose the subsidiary traceless condition $ {\cal K}_{\cdot \cdot   \gamma}=0$,  this field can be associated with the $j=p=2$  UIR of the discrete series of de Sitter group.
This is exactly the massless gauge field ${\cal K}_{\alpha\beta\gamma}$.

There exist another gauge field {\it i.e.}  the massless vector-spinor gauge fields $\Psi_\alpha$.
This gauge potential is a spinor and also a Grassmannian function, satisfying the anti-commutation relations. The corresponding infinitesimal generators $({\cal Q}_i)$ must be a spinor in the ambient space formalism and must satisfy the anti-commutation relations. Nevertheless, such algebra would not be closed, since the anti-commutation of two spinor generators become a tensor generator. In this case, in order to obtaining a closed super-algebra, the Grasmanian generators must be coupled with the generators of the dS group $L_{\alpha\beta}$ \cite{parota}. The $N=1$ super-algebra in the dS ambient space formalism is already established \cite{parota}:
$$ \{{\cal Q}_i,{\cal Q}_j\}=\left(S^{(\frac{1}{2})}_{\alpha\beta}\gamma^4
\gamma^2\right)_{ij}L^{\alpha\beta},\;\;
[{\cal Q}_i,L_{\alpha\beta}]=\left(S^{(\frac{1}{2})}_{\alpha\beta}{\cal Q} \right)_i, \;\; [\tilde
{\cal Q}_i,L_{\alpha\beta}]=-\left(\tilde {\cal Q} S^{(\frac{1}{2})}_{\alpha\beta}\right)_i,$$
\b \label{n1ssa}
[L_{\alpha\beta}, L_{\gamma\delta}] =
-i(\eta_{\alpha\gamma}L_{\beta\delta}+\eta_{\beta\delta}
L_{\alpha\gamma}-\eta_{\alpha\delta}L_{\beta\gamma}-\eta_{\beta\gamma}
L_{\alpha\delta}),\e
where $\tilde{{\cal Q}}_i=\left({\cal Q}^t \gamma^4
C\right)_i=\bar{{\cal Q}}_i$, and $Q^t$ is the transpose of ${\cal Q}$. The charge conjugation $C=\gamma^2\gamma^4$ was obtained in ambient space formalism \cite{morrota}. Now we can simply generalized the above discussion to the super gauge field $\Psi_\alpha$. The algebra (\ref{yma}) is replaced with the following super algebra:
 $$[X_A,X_B\}=f_{AB}^{\;\;\;\;\;C}X_C, \; A,B,C=1,...,n.$$
In this case we have $n$ gauge vector or vector-spinor fields $\Phi_\alpha^{\;\;A}$ . Since these gauge fields exist on the dS hyperboloid, they must be transverse with respect to the first index $x\cdot \Phi_{.}^{\;\;A}\equiv x^\alpha\Phi_{\alpha}^{\;\;A}=0$. In the ambient space notation the gauge-covariant derivative can be defined as
$$ D^{\Phi}_\beta=\nabla^\top_\beta+i \Phi_\beta^{\;\;A} X_A,$$
where $X_A\equiv (L_{\alpha\beta},{\cal Q}_i)$ and $\Phi_{\alpha}^{\;\;A}\equiv(K_\alpha^{\;\;a},\Psi_\alpha^{\;\; i}) $. In this case the curvature ${\cal R}$ is defined by:
$$ {\cal R}(D^{\Phi}_\alpha,D^{\Phi}_\beta)=-\left[D^{\Phi}_\alpha,D^{\Phi}_\beta\right\}=R_{\alpha\beta}^{\;\;\;\; A}X_A.$$
The gauge invariant action or Lagrangian in the dS background for the gauge potential $ \Phi_\beta^{\;\;A}$ can be written in the following form \cite{ta1403,mama,van,frts}:
\begin{equation}
\label{actionss} S[\Phi]=\int d\mu(x) R_{\alpha\beta}^{\;\;\;\; A}Q_{AB}R^{\alpha\beta B},\end{equation}
where $Q_{AB}$ is a numerical constant. For the super-gauge field $\Psi_\alpha$ the action was presented in \cite{paenta}.

The gauge invariant field equation is then obtained from the above action. $\Phi_\alpha^{\;\;A}$ can be set as  the vector-spinor gauge field $\Psi_\alpha$ and the gauge potential ${\cal K}_{\alpha\beta\gamma}$. In the framework of the gauge theory, since the vector-spinor gauge field $\Psi_\alpha$ must couple with the the gauge potential ${\cal K}_{\alpha\beta\gamma}$, the gravitational field could be composed by three elements; the dS background, the gravitational waves ${\cal K}_{\alpha\beta\gamma}$, and $\Psi_\alpha$. The action (\ref{actionss}) is invariant under the super-gauge group of the Lie $N=1$ super-symmetry dS algebra (\ref{n1ssa}).

\section{Conclusion}

In this paper we conclude that the ambient space formalism allows us to solve the some problems of QFT in dS universe. The gravitational waves and their supersymmetry partner can be considered as two elementary fields ${\cal K}_{\alpha\beta\gamma}$ and $\Psi_\alpha$, which are free of any infra-red divergence. The results are: 1) the infra-red divergence does not appear in this formalism for the linear gravity, 2) the QFT is constructed on Bunch-Davies vacuum state and it is unitary, 3) de Sitter super-algebra can be closed for odd $N$, and 4) the action of dS super-gravity  (\ref{actionss})  is real.

\vspace{0.3cm} 
\noindent {\bf{Acknowlegements}}:  We are grateful to Edouard Brezin,  Jean-Pierre Gazeau, Eric Huguet, Jean Iliopoulos, Richard Kerner, Salah Mehdi, Jacques Renaud and Shariar Rouhani for helpful discussion on this work.

\end{document}